 [2005/06/22 5.2/AAS markup document class]%
\documentclass[preprint]{aastex}%
\usepackage{graphicx}
\makeindex

\makeatletter 

\let\o@verbatim\verbatim
\def\verbatim{%
  \ifhmode\unskip\par\fi
  \ifx\@currsize\normalsize
     \small
  \fi
  \o@verbatim
}

\renewcommand \verbatim@font {%
  \normalfont \ttfamily
  \catcode`\<=\active
  \catcode`\>=\active
}

\RequirePackage{shortvrb}
\MakeShortVerb{\|}

\begingroup
  \catcode`\<=\active
  \catcode`\>=\active
  \gdef<{\@ifnextchar<\@lt\@meta}
  \gdef>{\@ifnextchar>\@gt\@gtr@err}
  \gdef\@meta#1>{\m{#1}}
  \gdef\@lt<{\char`\<}
  \gdef\@gt>{\char`\>}
\endgroup
\def\@gtr@err{%
   \ClassError{ltxguide}{%
      Isolated \protect>%
   }{%
      In this document class, \protect<...\protect>
      is used to indicate a parameter.\MessageBreak
      I've just found a \protect> on its own.
      Perhaps you meant to type \protect>\protect>?
   }%
}
\def\verbatim@nolig@list{\do\`\do\,\do\'\do\-}

\newcommand{\m}[1]{\mbox{$\langle$\it #1\/$\rangle$}}

\def\cmd#1{\cs{\expandafter\cmd@to@cs\string#1}}
\def\cmd@to@cs#1#2{\char\number`#2\relax}
\DeclareRobustCommand\cs[1]{\texttt{\char`\\#1}}
\def\GetFileInfo#1{%
  \def\filename{#1}%
  \def\@tempb##1 ##2 ##3\relax##4\relax{%
    \def\filedate{##1}%
    \def\fileversion{##2}%
    \def\fileinfo{##3}}%
  \edef\@tempa{\csname ver@#1\endcsname}%
  \expandafter\@tempb\@tempa\relax? ? \relax\relax}

\begin{document}

\makeatother
\title{Ganalyzer: A tool for automatic galaxy image analysis}
\author{Lior Shamir}
\affil{Dept. of Comp. Sci., Lawrence Tech U. \\ 21000 W Ten Mile Rd., Southfield, MI 48075}
\email{lshamir@mtu.edu}

\begin{abstract}
We describe Ganalyzer, a model-based tool that can automatically analyze and classify galaxy images. Ganalyzer works by separating the galaxy pixels from the background pixels, finding the center and radius of the galaxy, generating the radial intensity plot, and then computing the slopes of the peaks detected in the radial intensity plot to measure the spirality of the galaxy and determine its morphological class. Unlike algorithms that are based on machine learning, Ganalyzer is based on measuring the spirality of the galaxy, a task that is difficult to perform manually, and in many cases can provide a more accurate analysis compared to manual observation. Ganalyzer is simple to use, and can be easily embedded into other image analysis applications. Another advantage is its speed, which allows it to analyze $\sim$10,000,000 galaxy images in five days using a standard modern desktop computer. These capabilities can make Ganalyzer a useful tool in analyzing large datasets of galaxy images collected by autonomous sky surveys such as SDSS, LSST or DES. The software is available for free download at http://vfacstaff.ltu.edu/lshamir/downloads/ganalyzer, and the data used in the experiment is available at http://vfacstaff.ltu.edu/lshamir/downloads/ganalyzer/GalaxyImages.zip.
\end{abstract}


\expandafter\GetFileInfo\expandafter{\jobname.tex}%

\maketitle
\section{Introduction}
\label{introduction}

Robotic telescopes that acquire large datasets of astronomical images have introduced the need for methods and tools that can automatically analyze astronomical images and turn these data into knowledge. One of the tasks that require automation is the morphological analysis of galaxy images, which is an essential tool in sky surveys such as SDSS \citep{Yor00} or the future LSST \citep{Tys02}, DES \citep{Abb05}, and the space-based TAUVEX galaxy survey \citep{Bro07}.

The first attempts to automatically classify galaxies were made by \cite{Mor58,Mor59}, and followed by the work of \cite{Kor96}, who tried to classify elliptical galaxies by their internal structures. Other studies used central concentration as an indicator that can determine the position of a galaxy on the Hubble sequence \citep{Doi93,Bri98, Shi01}, or the central concentration and asymmetry of galaxian light \citep{Abr96}. Another approach to galaxy image analysis is the parametric approach, used by tools such as GIM2D \citep{Pen02} and GALFIT \citep{Sim98}, which can be wrapped by the GALAPAGOS script to improve its performance \citep{Hau07}.

Later attempts to perform automatic morphological classification of galaxies include the Gini coefficient method \citep{Abr03}, and the CAS method \citep{Con03}. However, the efficacy of these methods for real-life galaxy morphological classification has been criticized \citep{Tho08}, and did not provide solid useful tools that can be used for galaxy morphological analysis \citep{Lin08}. This led to the contention that practical classification of large datasets of galaxy images should be carried out by humans \citep{Lin08}.

A significant improvement was introduced by the application of machine learning approaches to the task of galaxy classification. \cite{Hue08,Hue09} used Support Vector Machine for galaxy classification and probability density estimation, and applied the method to SDSS DR7 \citep{Hue11}. \cite{Bal04,Bal08} achieved good results by utilizing an Artificial Neural Network. Recent studies also showed significant improvement in the accuracy of automatic classification of galaxy images used by the Galaxy Zoo project, demonstrating accuracy of $\sim$90\% for the classification of the three primary morphological types (spiral, elliptical, and edge-on), and $\sim$95\% accuracy when classifying spiral and elliptical galaxies \citep{Sha09,Ban10}. While showing good classification accuracy, these machine learning methods require a step of training, and normally do not provide useful information about the galaxy other than its morphological type. Here we describe Ganalyzer, which is a fast and easy to use model-based tool that measures the ellipticity and spirality of galaxies. In Section~\ref{method} the image analysis method is described, Section~\ref{results} discusses the performance evaluation and experimental results, and Section~\ref{ganalyzer} provides an introduction to using the Ganalyzer command line utility.

\section{Morphological analysis method}
\label{method}

\subsection{Finding the galaxy center, ellipticity, and position angle}
\label{prop}
\begin{figure}[p]
\plotone{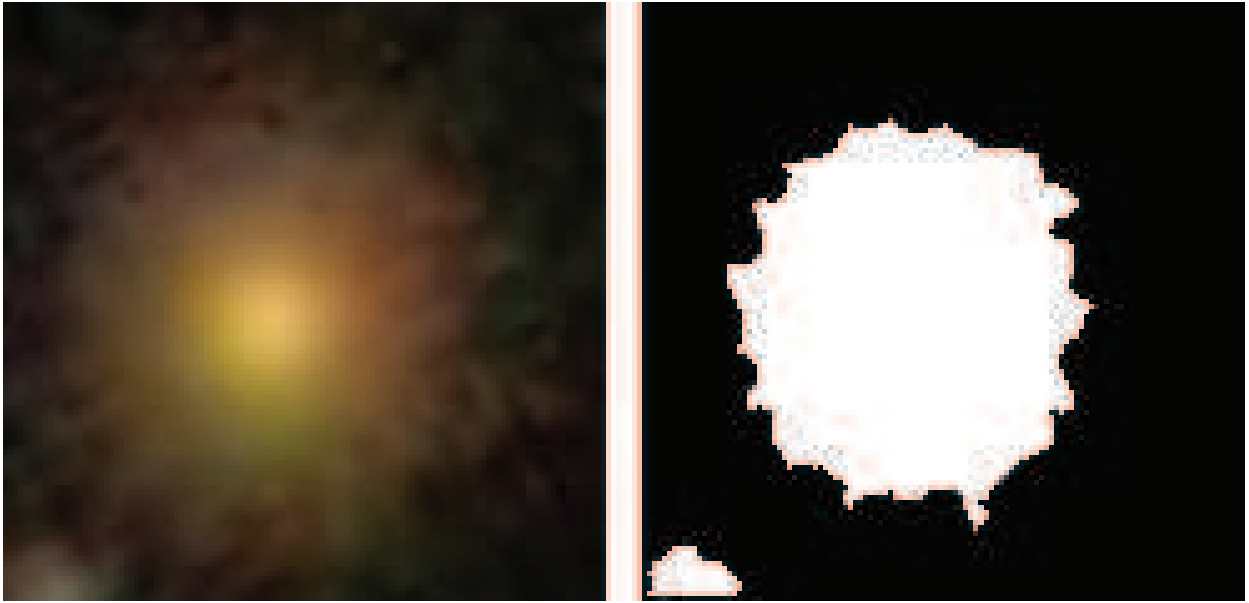}
\caption{A galaxy image (left) and the pixels detected as foreground using the Otsu method}
\label{otsu_image}
\end{figure}

The first step of Ganalyzer is detecting the objects in the image and extracting basic information about each object such as the center, ellipticity and position angle, as done by object detection methods such as SExtractor \citep{Ber96}. This goal is achieved by first separating the objects from the background by applying the Otsu threshold \citep{Ots79}, which is a widely used method for determining the graylevel threshold that separates between the foreground and the background pixels. Figure~\ref{otsu_image} shows an original galaxy image taken from {\it Galaxy Zoo} and the foreground galaxy pixels detected using the Otsu method.

\placefigure{otsu_image}

Once foreground pixels are separated from the background pixels, all eight-connected objects that their surface size is larger than 1000 pixels are detected. Detecting objects in the image allows ganalyzer to analyze images in which more than one object is present, and the 1000 threshold is used to reject small foreground objects (surface size smaller than 1000 pixels) that are present in the image but are too small to provide an interpretable morphological structure.

After the objects are detected, each object is assigned with its pixel mass center, computed as the image coordinates $(v,w)$ such that the number of pixels $(x,y)$ where $x<v$ equals the number of pixels $(x,y)$ where $x>v$ \citep{wndchrm}. Similarly, the $w$ coordinate is computed such that the number of pixels where $y<w$ equals the number of pixels such that $y>w$. Then, the galaxy center $(O_x,O_y)$ is determined as the center of the 5$\times$5 shifted window that has the highest median value and its distance from the pixel mass center of the object is smaller than $0.1/ \sqrt{S \over \pi}$, where {\it S} is the surface size of the object in pixels. This simple and fast method accurately detected the center of the galaxy in all 525 galaxy images tested in this study.

After the galaxy center is found, the radius of the galaxy is determined by the maximal distance between the object center and any foreground pixel. The major axis of the galaxy is determined as the longest possible line that passes through the center, and the minor axis is determined by the length of the line that passes through the center in 90$^o$ to the major axis. The ellipticity of the galaxy is then determined by the minor axis of the galaxy divided by its major axis. In addition to the ellipticity, ganalyzer also computes the position angle of the galaxy.

\subsection{Detecting spirality}
 
The basic element used in this study for measuring spirality is the radial intensity plot. The radial intensity plot is a 360$\times$35 image, such that the value of the pixel $(x,y)$ is the median value of the 5$\times$5 windows around the pixel at image coordinates $(O_x+sin(\theta) \cdot r,O_y-cos(\theta)\cdot r)$ in the galaxy image, where $\theta$ is the polar angle (in degrees) and r is a radial distance. Intuitively, the radial intensity plot is an image of the radial intensities at different distances from the galaxy center. Each horizontal line in the radial intensity plot is then smoothed using a median filter with a span of 50 pixels. If the radius of the galaxy is greater than 100 pixels, the radial intensity plot is computed after downscaling the object such that the radius is set to 100. This practice helps to analyze high-resolution images in which star forming regions or substructures in the spirals can affect the detection of the arms.

\begin{figure}[p]
\plotone{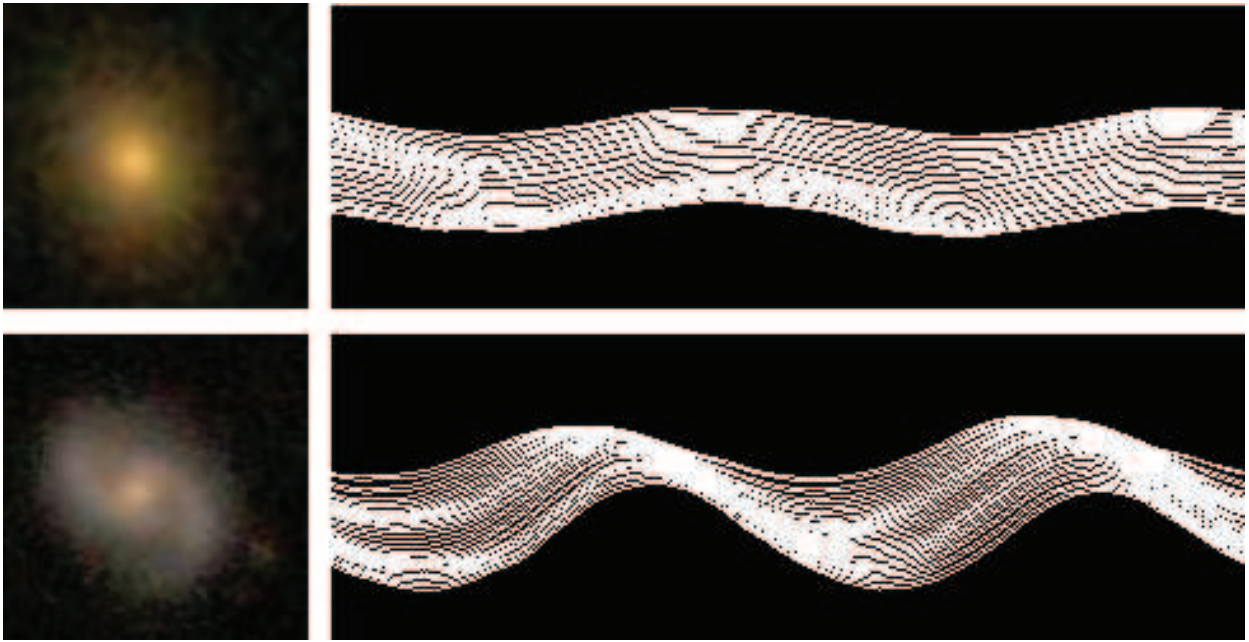}
\caption{Galaxy images (left) and the transformation of the radial intensity plots such that the Y axis is the intensity and the X axis is the polar angle}
\label{radial}
\end{figure}

Figure~\ref{radial} shows the original galaxy image and a transformation of the radial intensity plot such that the Y axis is the intensity and the X axis is the polar angle. As the figure shows, in an image of a spiral galaxy the peaks are expected to shift due to the spirality of the arms, while if the galaxy is not spiral the peaks are expected to align in a near-straight line.

\placefigure{radial}

\begin{figure}[p]
\plotone{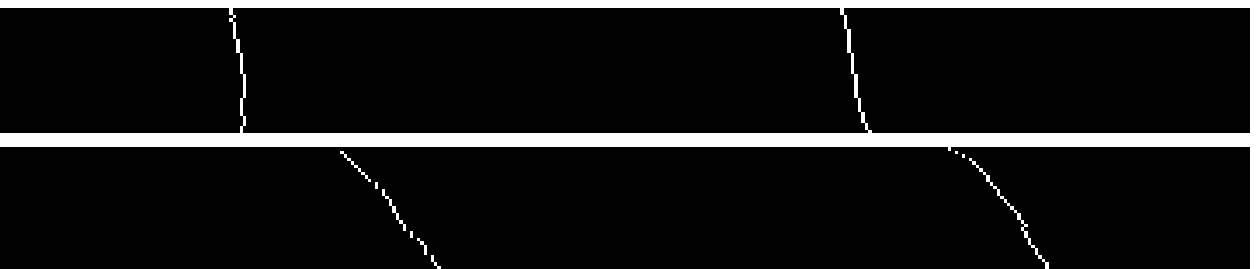}
\caption{The peaks detected in the radial intensity plots of the elliptical (up) and spiral galaxies of Figure~\ref{radial} }
\label{peaks}
\end{figure}

To use the shift of the peaks in the radial intensity plot for detecting spirality, the peaks are first detected using an automatic peak detection algorithm \citep{Mor00}, with the parameters $\sigma$=10, {\it threshold}=0.05, and {\it iterations}=1, as described in \citep{Mor00}. Figure~\ref{peaks} shows the peaks detected in the galaxy images of Figure~\ref{radial}, such that the radial distance is between 0.4r to 0.75r, where {\it r} is the radius of the galaxy described in Section~\ref{prop}.

\placefigure{peaks}

Once the peaks are detected, a linear regression is used to determine the slope of each of the two groups of peaks that have the highest number of detected peaks.  For instance, in the galaxies of Figure~\ref{peaks} the peaks of the spiral galaxy are organized in two lines with slopes of $\sim$0.74 and $\sim$0.81, while the slopes of the peaks of the elliptical galaxy are $\sim$0.22 and $\sim$0.1. The slopes of the arm reflect the level of spirality of the examined galaxy. To avoid the effect of local variations that can lead to peaks such as stars or satellite galaxies, only groups that have 20 or more peaks are included in the analysis.

If just one arm is detected, a galaxy is considered spiral if the absolute value of the slope of the arm is greater than 0.35. If more than one arm is detected, the analysis is based on the two arms with the largest number of peaks. If both arms have slopes greater than 0.5, or if one of the arms has a slope greater than 0.7, then the galaxy is considered by Ganalyzer as spiral. If the standard deviation of one of the arms is smaller than 2, then a slope greater than 0.35 in this arm will also be considered by Ganalyzer as an indication of a spiral galaxy. These rules were determined experimentally by comparing the analysis of the slopes to the manual galaxy classification performed by the author using the galaxy image datasets used in \citep{Sha09}.

The slope of the arm of a spiral galaxy can peak at different distances from the center in different galaxies. Therefore, the peaks are detected in four different ranges of distances from the center: 0.1r to 0.45r, 0.2r to 0.55r, 0.3r to 0.65r, and 0.4r to 0.75r, such that {\it r} is the radius of the galaxy described in Section~\ref{prop}. If the slopes of the peaks detected in any of these ranges meet the criteria described above, then the galaxy is determined to be spiral.

Ganalyzer also detects the presence of bars, which is performed by analyzing the vertical lines in the radial intensity plot generated for distances 0.5r to 1.0r. While the intensity is normally expected to decrease when moving away from the galaxy center, if bars exist it is expected that the intensities will increase at around the distance of the bar from the center. Therefore, if 50\% or more of the vertical lines of the radial intensity plot show an intensity increase the galaxy is determined to have bars.

If no spirality is detected, the galaxy is determined to be edge-on if the ellipticity described in Section~\ref{prop} is greater than 0.8. Otherwise, the galaxy is considered elliptical. It should be noted that Ganalyzer outputs the ellipticity value, which can be more informative than the distinct morphological class.

\section{Experimental Results}
\label{results}

Ganalyzer was tested using a dataset of small galaxy images taken from the {\it Galaxy Zoo} web site \citep{Lin08}, and was previously used for developing a machine learning-based galaxy image classification method \citep{Sha09}. The first dataset contains 225 images classified manually by the author as spiral, 225 images classified as elliptical, and 75 galaxy images classified as edge-on galaxies. All images were color images, and did not contain {\it Galaxy Zoo} monochrome images that were collected for the {\it Galaxy Zoo} bias study \citep{Lin08}. The dataset is available for free download at  http://vfacstaff.ltu.edu/lshamir/downloads/ganalyzer/GalaxyImages.zip.

Among the 525 galaxy images, 466 were classified correctly, which is $\sim$89\% of accuracy where the gold standard is the manual classification performed by the author. Table~\ref{confusion_matrix} shows the confusion matrix of the classification.

\placetable{confusion_matrix}

\begin{table}[p]
\footnotesize
\begin{center}
\caption{Confusion matrix of the galaxy classification}
\label{confusion_matrix}
\begin{tabular}{@{}lccc@{}} 
 & Spiral & Elliptical & Edge-on \\
\hline
Spiral &  206 & 19 & 0 \\
Elliptical & 34 & 191 & 0 \\
Edge-on & 3 & 3 & 69 \\
\end{tabular}
\end{center}
\end{table}

\begin{figure}[p]
\epsscale{0.25}
\plotone{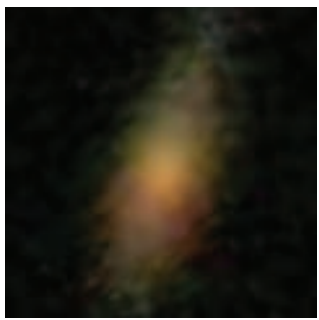}
\caption{A galaxy image that was classified manually as edge-on and as elliptical by Ganalyzer}
\label{edge_on_elliptical}
\end{figure}

While the classification accuracy of $\sim$89\% is less than perfect, it should be noted that manual classification is subjective, and might also not provide a fully reliable ``gold standard" due to the many in-between cases. For instance, the galaxy in Figure~\ref{edge_on_elliptical} was classified manually as edge-on, but Ganalyzer classified it as elliptical. In this case the galaxy seemed to the person classifying it long and narrow enough to be classified as an edge-on galaxy, while Ganalyzer assigned it with a relatively high ellipticity value of $\sim$0.62 but classified it as elliptical.

\placefigure{edge_on_elliptical}

While in some cases disagreements between Ganalyzer and manual classification can be due to in-between cases, in other cases Ganalyzer can detect features that are difficult to notice with casual observation of a galaxy image using the unaided eye. Table~\ref{spiral_elliptical} shows galaxy images that were classified manually as elliptical galaxies, but Ganalyzer classified them as Spiral. 

\placetable{spiral_elliptical}

\begin{table}[p]
\footnotesize
\begin{center}
\caption{Galaxy images that were classified as elliptical manually and as spiral by Ganalyzer.}
\label{spiral_elliptical}
\begin{tabular}{@{}lcc@{}}  
Galaxy Zoo ID & Image & Radial Intensity Plot \\
\hline
587736945741201714 & \includegraphics[scale=0.6]{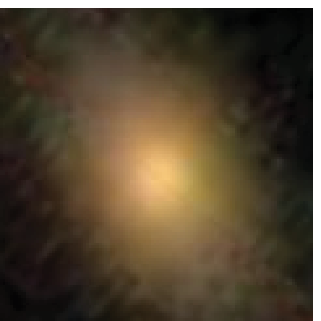} & \includegraphics[scale=0.6]{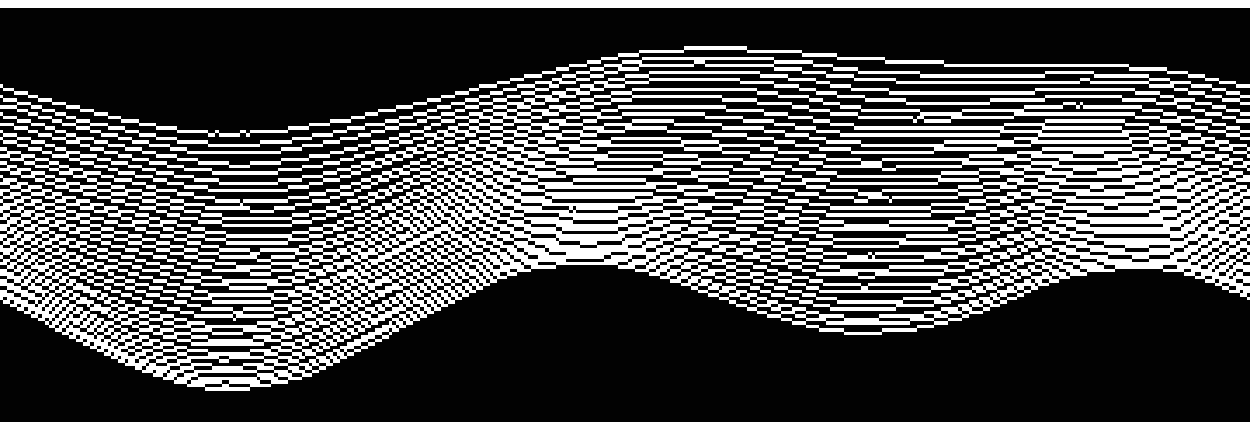} \\
587737810113396966 & \includegraphics[scale=0.6]{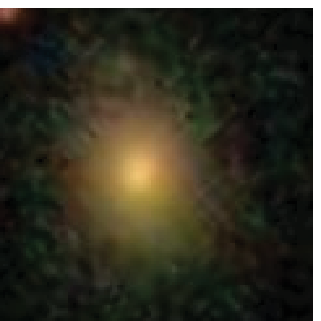} & \includegraphics[scale=0.6]{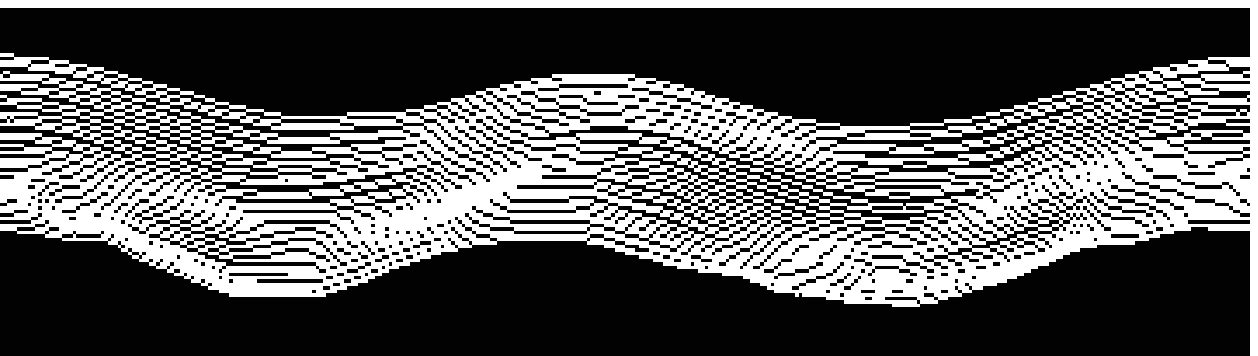} \\
587733397576745343 & \includegraphics[scale=0.6]{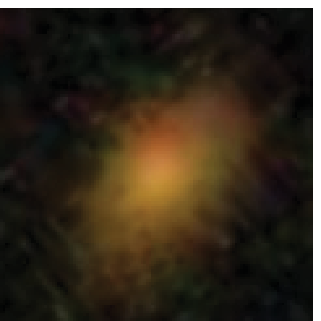} & \includegraphics[scale=0.6]{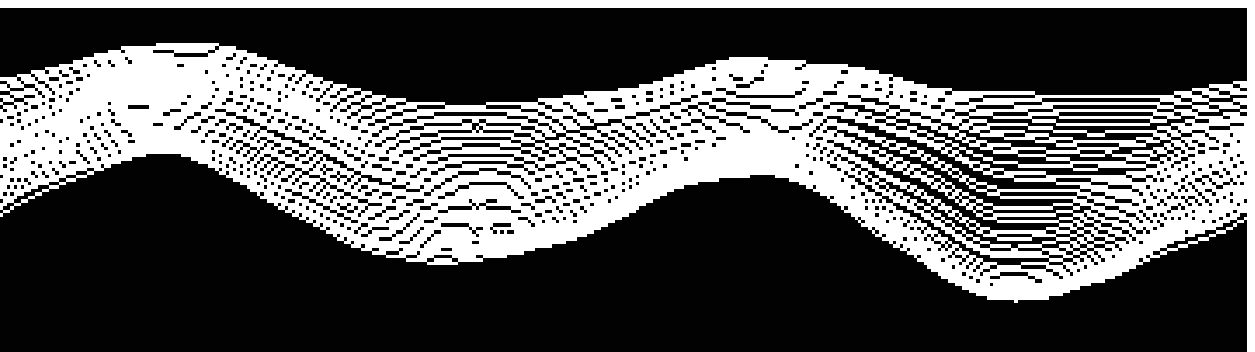} \\
587735348010156087 & \includegraphics[scale=0.6]{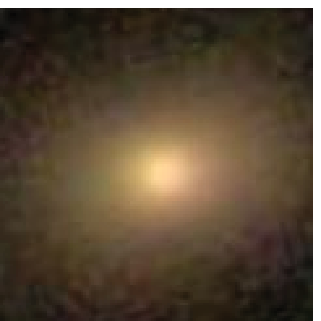} & \includegraphics[scale=0.6]{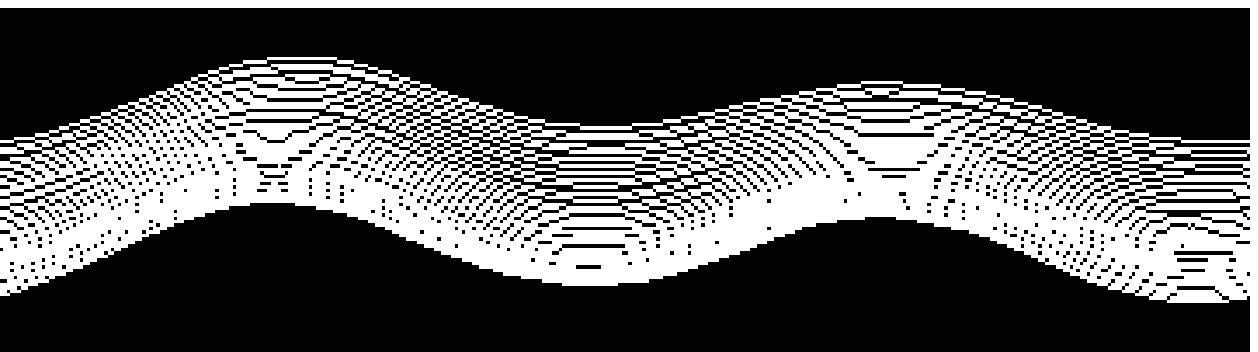} \\
587737826749841683 & \includegraphics[scale=0.6]{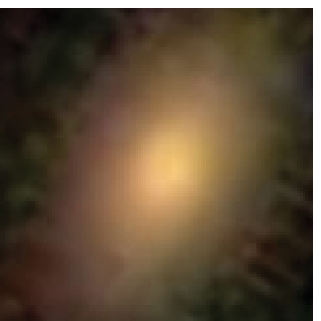} & \includegraphics[scale=0.6]{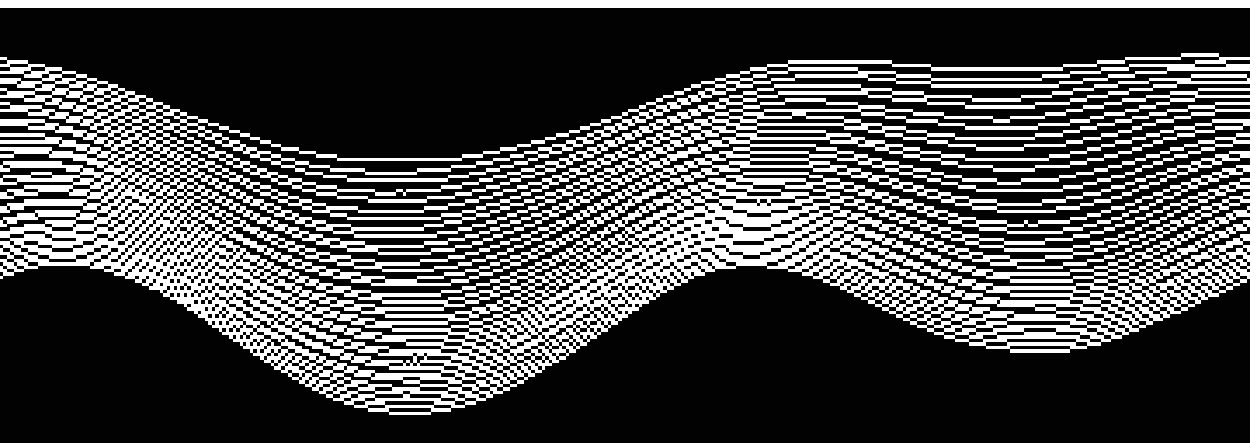} \\
587742189916717258 & \includegraphics[scale=0.6]{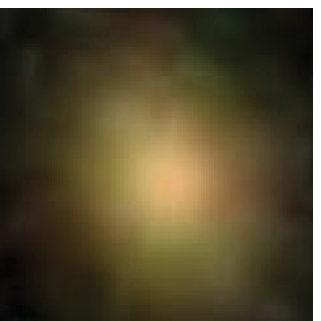} & \includegraphics[scale=0.6]{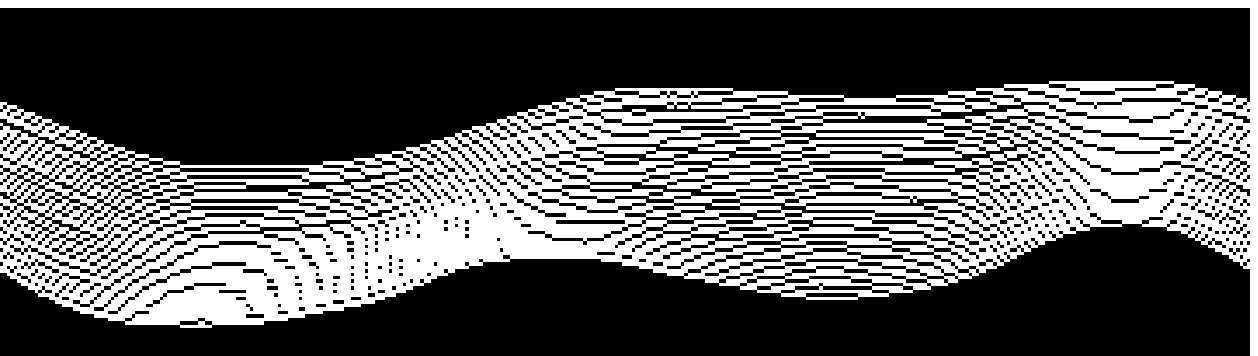} \\
587742551753883805 & \includegraphics[scale=0.6]{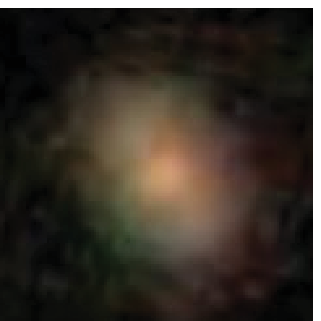} & \includegraphics[scale=0.6]{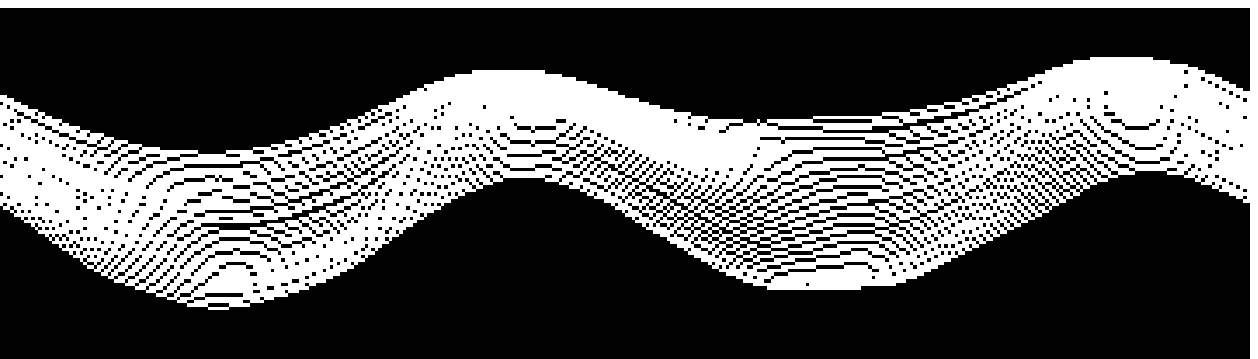} \\
588007004703621329 & \includegraphics[scale=0.6]{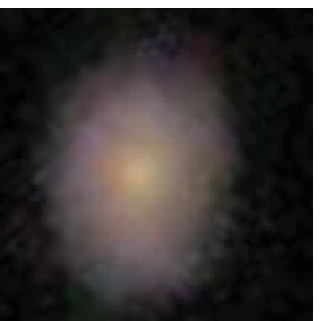} & \includegraphics[scale=0.6]{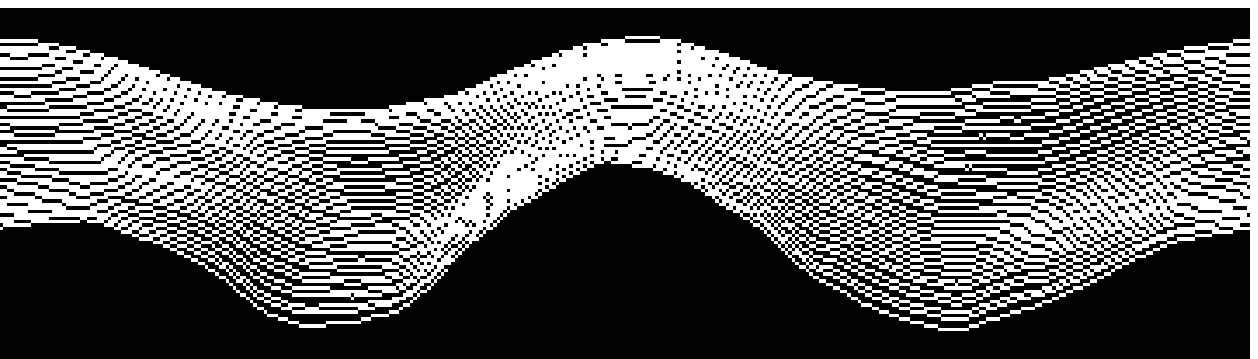} \\
588010360157831362 & \includegraphics[scale=0.6]{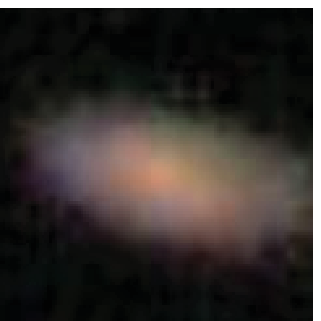} & \includegraphics[scale=0.6]{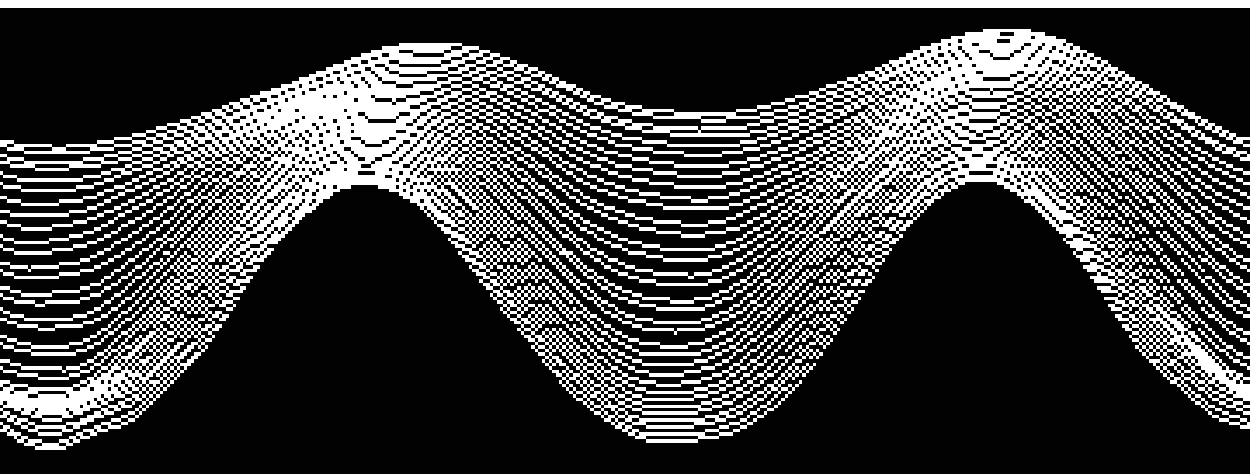} \\

\end{tabular}
\end{center}
\end{table}

As the table shows, the radial intensity plots of these galaxies indicate that some of the peaks shift as the distance from the center changes, which might indicate that these galaxies feature spirality. This spirality might be difficult to detect using the unaided eye, but can be detected more easily by Ganalyzer using the radial intensity plots. As Table~\ref{confusion_matrix} shows, most of the disagreements between the manual classification and Ganalyzer were in galaxies that were classified manually as elliptical while Ganalyzer classified them as spiral. However, in some cases galaxies that were classified manually as spiral were classified by Ganalyzer as elliptical.

\placetable{elliptical_spiral}

\begin{table}[p]
\footnotesize
\begin{center}
\caption{Galaxy images that were classified as spiral manually but were classified as elliptical by Ganalyzer.}
\label{elliptical_spiral}
\begin{tabular}{@{}lcc@{}}
Galaxy Zoo ID & Image & Radial Intensity Plot \\
\hline
587729229300039765 & \includegraphics[scale=0.6]{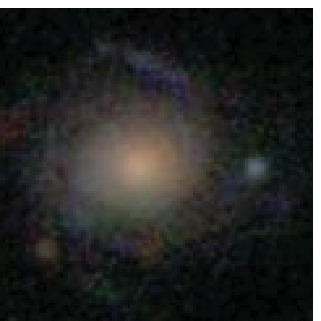} & \includegraphics[scale=0.6]{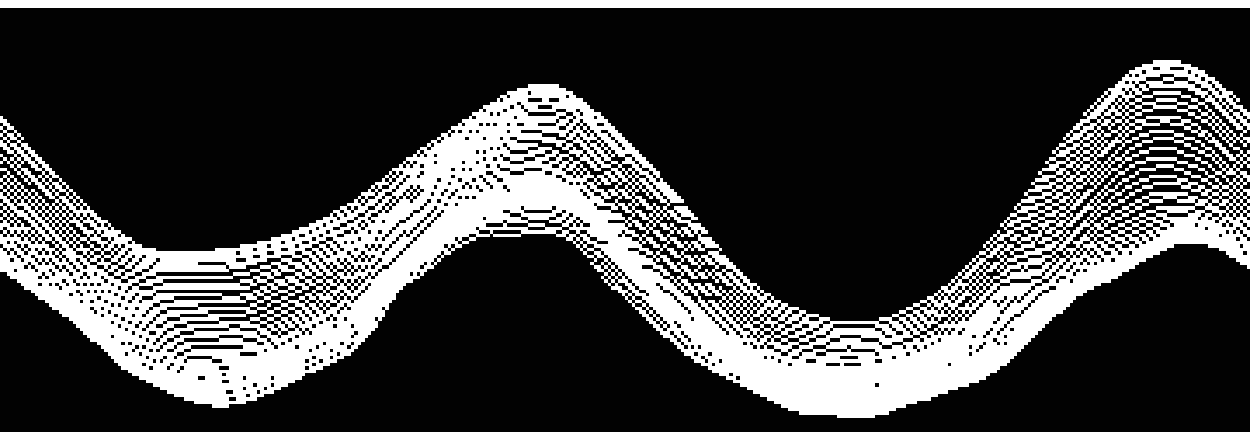} \\

587732483292790857 & \includegraphics[scale=0.6]{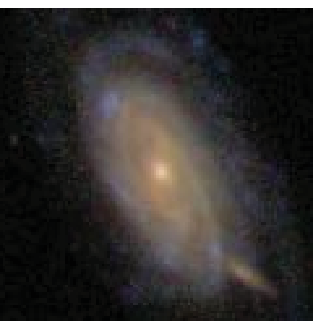} & \includegraphics[scale=0.6]{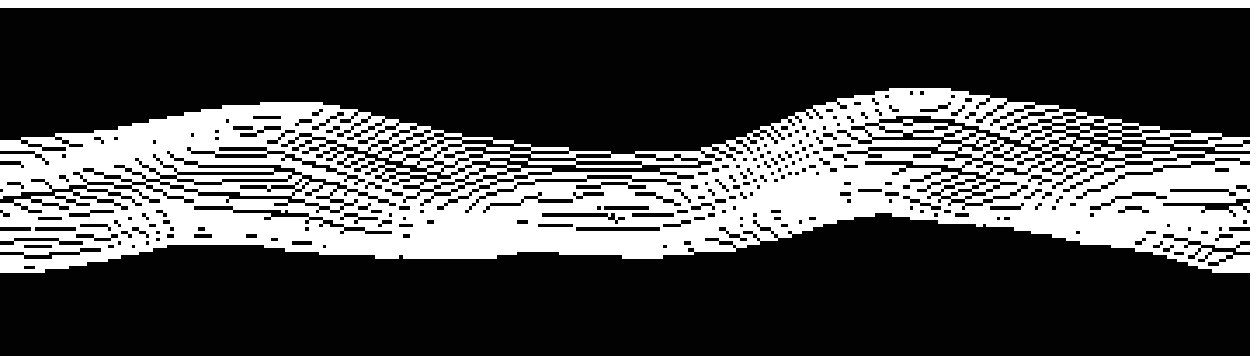} \\

587738067262374254 & \includegraphics[scale=0.6]{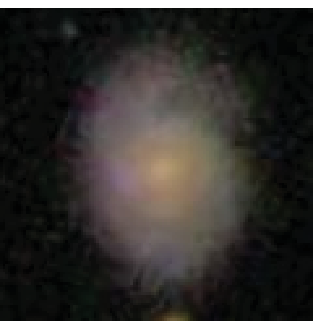} & \includegraphics[scale=0.6]{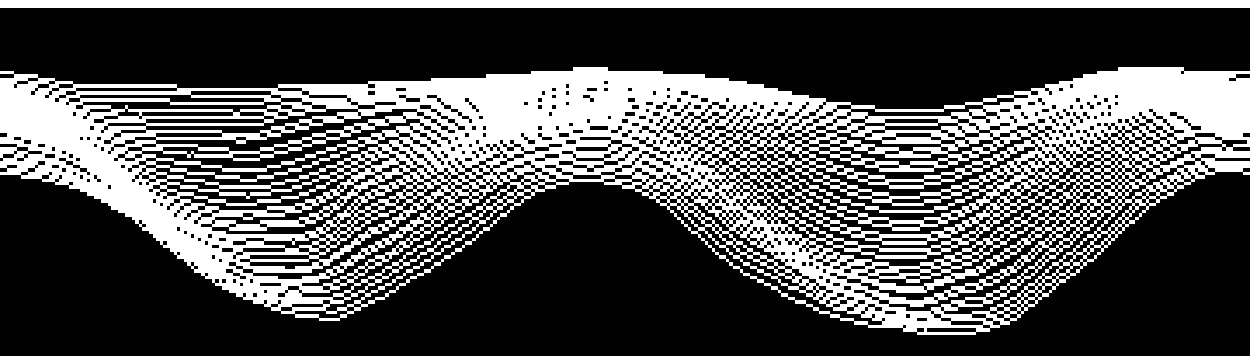} \\

587739457225949323 & \includegraphics[scale=0.6]{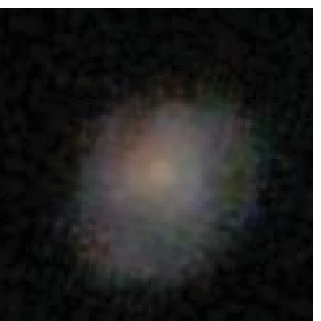} & \includegraphics[scale=0.6]{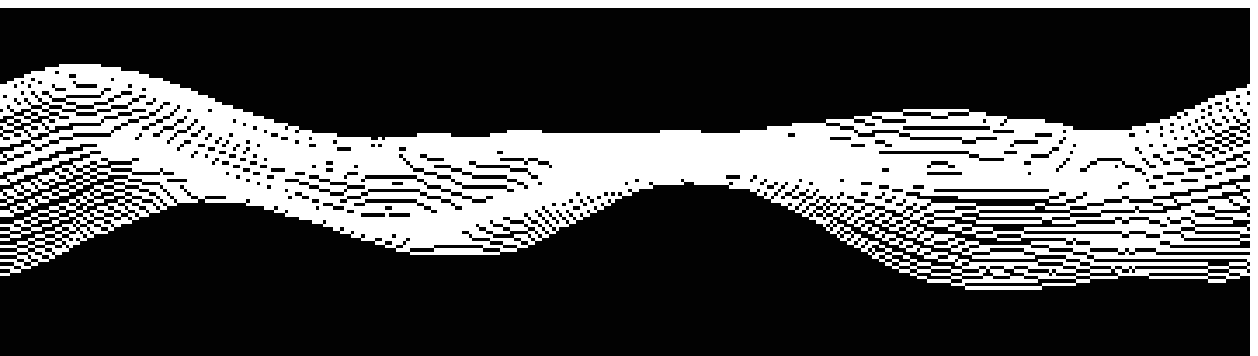} \\

587739707406024797 & \includegraphics[scale=0.6]{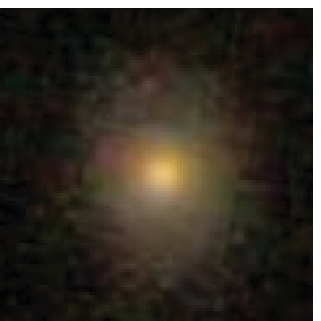} & \includegraphics[scale=0.6]{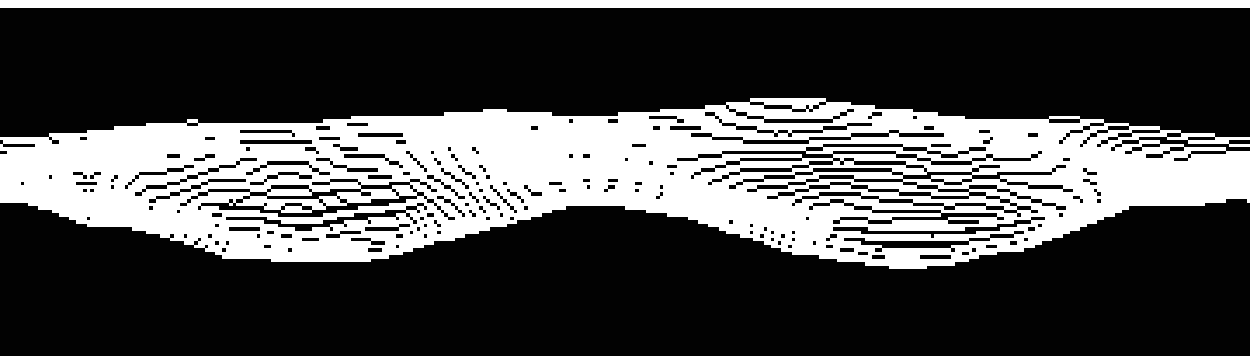} \\

587739720832647252 & \includegraphics[scale=0.6]{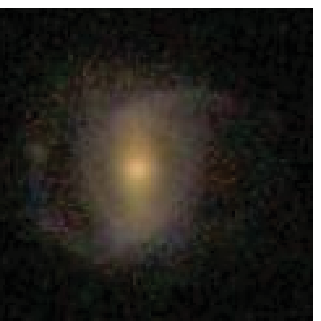} & \includegraphics[scale=0.6]{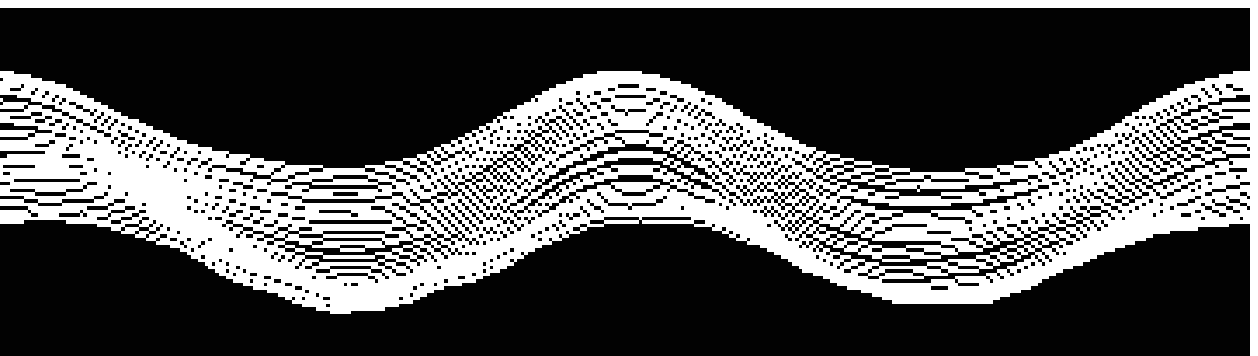} \\

587741600950452406 & \includegraphics[scale=0.6]{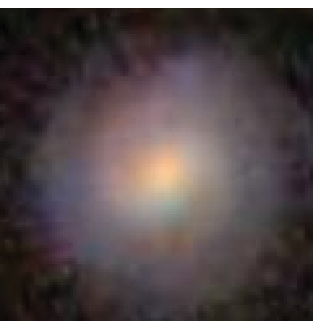} & \includegraphics[scale=0.6]{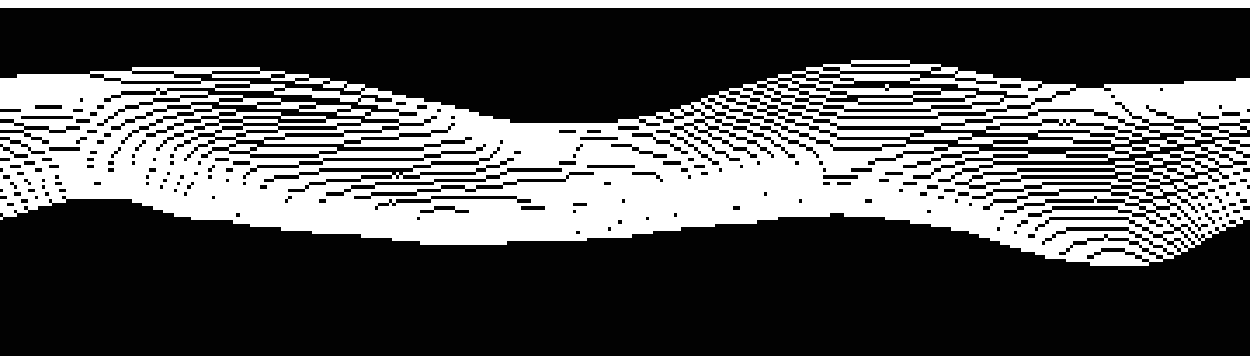} \\

587741709947371602 & \includegraphics[scale=0.6]{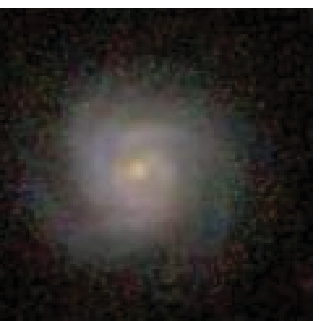} & \includegraphics[scale=0.6]{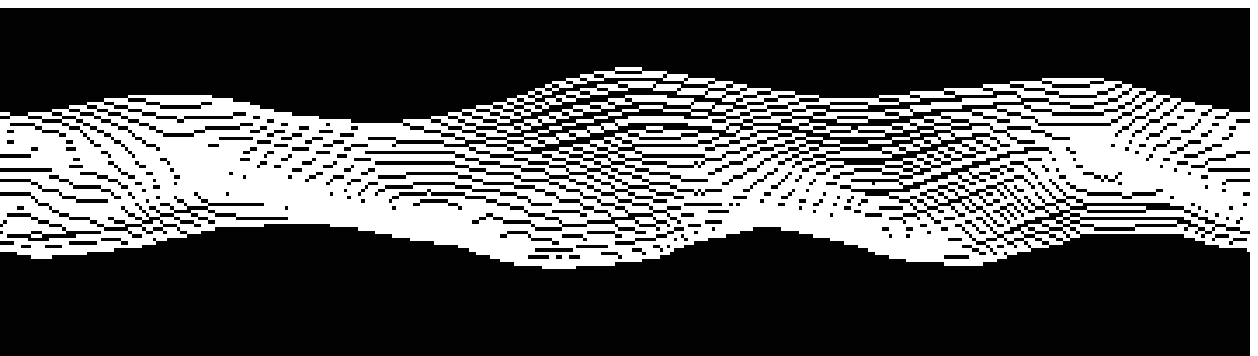} \\

587741710494662777 & \includegraphics[scale=0.6]{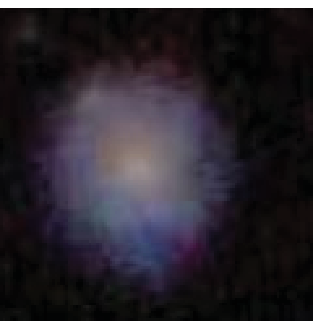} & \includegraphics[scale=0.6]{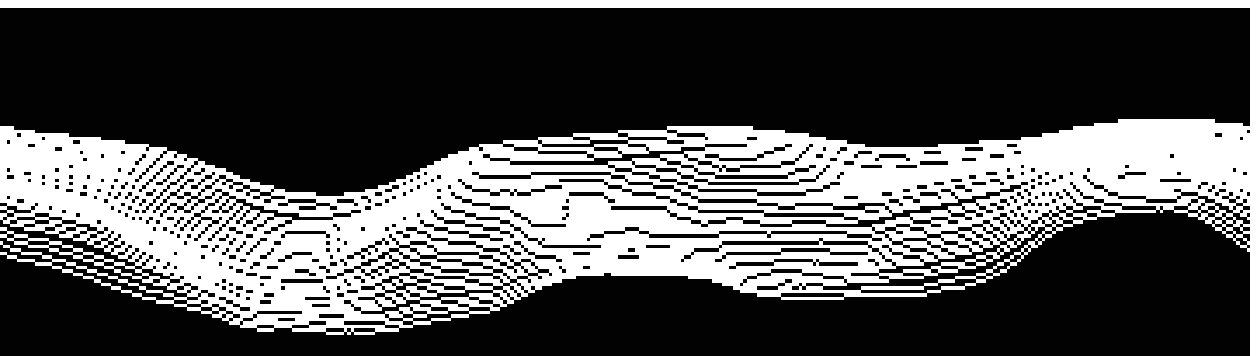} \\

\end{tabular}
\end{center}
\end{table}

Table~\ref{elliptical_spiral} shows spiral galaxies that were classified incorrectly. As the table shows, these galaxies were clearly classified incorrectly by Ganalyzer. While the radial intensity plots show that some of the peaks shift as the distance from the galaxy center changes, in some cases the peaks are not always detected correctly, and improving the peak detection used in this study \citep{Mor00} might improve the performance of the galaxy classification. Another limitation of Ganalyzer is that the analysis is dependent on the arms, and therefore if the resolution of the image is too low and the arms cannot be seen the galaxy might not be analyzed correctly. 

To test ganalyzer with a larger set of galaxy images, another experiment was performed with a galaxy dataset of 6209 galaxy images classified as spiral and 2316 galaxy images classified as elliptical by Galaxy Zoo participants \citep{Lin11}. The experiment showed that ganalyzer was in agreement with the manual classification in $\sim$86\% of the cases.

An important advantage of Ganalyzer is its low computational complexity, which allows it to process very many images using relatively modest computing resources. For instance, the galaxy image dataset of 525 images used in this study was processed in $\sim$170 seconds using a single core of Intel core-i7 quad core processor. Therefore, by using eight cores a standard desktop computer can process $\sim$10,000,000 images in just five days.

\section{Using Ganalyzer}
\label{ganalyzer}

The Ganalyzer tool is a simple Windows command-line utility that receives a path to a galaxy image file, and prints the analysis results to the standard output. For instance, the following command line returns the morphological class, as well as the ellipticity, position angle, surface size (pixels), radius, image coordinates of the center, and the slopes of the shifts of the peaks detected in the image ``galaxy.tif". \newline

C:\textbackslash $>$ ganalyzer c:\textbackslash path\textbackslash to\textbackslash galaxy.tif   \newline \newline
For instance, the output of Ganalyzer when applied on the spiral galaxy of Figure~\ref{radial} is: \newline \newline
Object 1: \newline
Center: (53,62)  \newline
Surface size (pixels): 3989  \newline
Radius (pixels): 45  \newline
Ellipticity: 0.295  \newline
Position angle: 143 degrees  \newline
slopes: 0.74 (stderr 1.28) 0.81 (stderr 0.80)  \newline
Galaxy type: Spiral  \newline

Currently, the supported file formats are TIFF, JPG, PPM, and BMP. In cases where the source images are in the FITS format, the images can be converted to lossless 8 or 16 bit TIFF format before being analyzed by Ganalyzer. Since Ganalyzer is used as a command line utility, it can be easily embedded into other applications and serve as a component in an astronomical pipeline processing system.

To allow a more informative analysis of the galaxy image, Ganalyzer can also output the radial intensity plots and the peaks. This can be done by using the ``-i" switch. For instance, the following command line can be used to generate the radial intensity plot and its transformation described in Figure~\ref{radial}, as well as the detected peaks as described in Figure~\ref{peaks}. \newline

C:\textbackslash $>$ ganalyzer -i c:\textbackslash path\textbackslash to\textbackslash galaxy.tif   \newline

When the ``i" switch is used, these images are created in the working directory, such that irp.tiff and irp\_radial.tiff are the radial intensity plot and the transformation described in Figure~\ref{radial}, and irp\_peaks.tiff is the image of the detected peaks as described in Figure~\ref{peaks}.

\section{Conclusion}
\label{conclusion}

The increasing availability of robotic telescopes that acquire large datasets of galaxy images has introduced the need for automatic methods for galaxy image analysis that can be practically used for analyzing these datasets. Ganalyzer is a fast and simple software tool that uses the radial intensity plots of galaxy images to measure the spirality and ellipticity of galaxies and classify galaxy images into three morphological classes of spiral, elliptical, and edge-on.

Ganalyzer is based on measuring the spirality of galaxies, and might not be optimal for detecting morphological features that are not directly related to the ellipticity and spirality. Therefore, ganalyzer might not excel in detecting galaxies that their unique morphology is not based on spirality such as S0, mergers, or peculiar galaxies.

The tool is used as a command-line utility, so that it can be embedded into other programs and serve as a component in a more comprehensive system of astronomical image pipeline processing. Since Ganalyzer is relatively quick, it can be practically used for analyzing very large datasets containing millions of galaxy images. Ganalyzer can be downloaded freely at:
http://vfacstaff.ltu.edu/lshamir/downloads/ganalyzer or from the Astrophysics Source Code Library at http://ascl.net.

\section{Acknowledgments}

I would like to thank Kevin Gravir for his assistance in this work, and the anonymous referee for his/her insightful and constructive comments. Funding for the SDSS and SDSS-II has been provided by the Alfred P. Sloan Foundation, the Participating Institutions, the National Science Foundation, the US Department of Energy, the National Aeronautics and Space Administration, the Japanese Monbukagakusho, the Max Planck Society, and the Higher Education Funding Council for England. The SDSS Web Site is http://www.sdss.org/. The SDSS is managed by the Astrophysical Research Consortium for the Participating Institutions. The Participating Institutions are the American Museum of Natural History, Astrophysical Institute Potsdam, University of Basel, University of Cambridge, Case Western Reserve University, University of Chicago, Drexel University, Fermilab, the Institute for Advanced Study, the Japan Participation Group, Johns Hopkins University, the Joint Institute for Nuclear Astrophysics, the Kavli Institute for Particle Astrophysics and Cosmology, the Korean Scientist Group, the Chinese Academy of Sciences (LAMOST), Los Alamos National Laboratory, the Max Planck Institute for Astronomy (MPIA), the Max Planck Institute for Astrophysics (MPA), New Mexico State University, Ohio State University, University of Pittsburgh, University of Portsmouth, Princeton University, the United States Naval Observatory and the University of Washington.


\begin{thebibliography}{}

\bibitem[Abbot(2006)]{Abb05}
Abbott, T., et al., 2006, {\it AIP Conference Proceedings}, 842, 989

\bibitem[Abraham et al.(1996)]{Abr96}
Abraham, R. G., Tanvir, N. R., Santiago, B. X., Ellis, R. S., Glazebrook, K., van den Bergh, S., 1996, \mnras, 279, L47

\bibitem[Abraham, Van Den Bergh \& Nair (2003)]{Abr03}
Abraham, R. G., Van Den Bergh, S., Nair, P., 2003, \apj, 588, 218

\bibitem[Ball et al.(2004)]{Bal04}
Ball, N. M., Loveday, J., Fukugita, M., Nakamura, O., Okamura, S., Brinkmann, J., Brunner, R. J., 2004, \mnras, 348, 1038

\bibitem[Ball et al.(2008)]{Bal08} 
Ball, N. M., Brunner, Robert J., Myers, A. D., Strand, N. E., Alberts, Stacey L., Tcheng, D.,  2008, \apj, 683, 12 

\bibitem[Banerji et al.(2010)]{Ban10}
Banerji, M., et al., 2010, \mnras, 406, 342

\bibitem[Bertin \& Arnouts(1996)]{Ber96}
Bertin, E., Arnouts, S. 1996, \aaps, 317, 393 

\bibitem[Brinchmann et al.(1998)]{Bri98}
Brinchmann, J., et al., 1998, \apj,  499, 112

\bibitem[Brosch \& Almoznino(2007)]{Bro07}
Brosch, N., Almoznino, E., {\it Bulletin of the Astronomical Society of India}, 35, 283

\bibitem[Conselice(2003)]{Con03}
Conselice, C.J., 2003, \apjs, 147, 1

\bibitem[Doi, Fukugita \& Okamura (1993)]{Doi93}
Doi, M., Fukugita, M., Okamura, S., 1993, \mnras, 264, 832

\bibitem[Haussler(2007)]{Hau07}
Haussler, B., McIntosh, D. H., Barden, M. et al., 2007, \apjs, 172, 615

\bibitem[Huertas-Company et al.(2008)]{Hue08}
Huertas-Company, M., Rouan, D., Tasca, L., Soucail, G., Le Fevre, O., 2008, \aap, 478, 971

\bibitem[Huertas-Company et al.(2009)]{Hue09} 
Huertas-Company, M., Tasca, L., Rouan, D., Pelat, D., Kneib, J. P., Le Favre, O., Capak, P., Kartaltepe, J., Koekemoer, A., McCracken, H. J., Salvato, M., Sanders, D. B., Willott, C., 2009, \aap, 497, 743

\bibitem[Huertas-Company et al.(2011)]{Hue11}
Huertas-Company, M., Aguerri, J. A. L., Bernardi, M., Mei, S., Sanchez Almeida, J., 2011, \aap, 525, 157 

\bibitem[Kormendy \& Bender(1996)]{Kor96}
Kormendy, J., Bender, R., 1996, \apj, 464, L119

\bibitem[Lintott et al.(2008)]{Lin08}
Lintott, C. J., et al. 2008, \mnras, 389, 1179

\bibitem[Lintott et al.(2011)]{Lin11} 
Lintott, C., Schawinski, K., Bamford, S., et al., \mnras, 2011, 410, 166 

\bibitem[Morhac et al.(2000)]{Mor00}
Morhac, M. et al., 2000, {\it Methods in Research Physics A}, 443, 108.

\bibitem[Morgan \& Mayall(1957)]{Mor58}
Morgan, W.W., Mayall, N. U., 1957, \pasp, 69, 291

\bibitem[Morgan \& Mayall(1969)]{Mor59}
Morgan, W.W., Osterbrock, D. E., 1969, \aj, 74, 515


\bibitem[Otsu(1979)]{Ots79}
Otsu, N., 1979, {\it IEEE Trans. Sys. Man. Cyber.} 9, 62

\bibitem[Peng et al.(2002)]{Pen02}
Peng, C. Y., Ho, L. C., Impey, C. D., Rix, H. W., 2002, \aj, 124, 266

\bibitem[Simard(1998)]{Sim98}
Simard, L., 1998, {\it ASP Conference Series}, 145, 108

\bibitem[Shamir et al.(2008)]{wndchrm}
Shamir, L., Orlov, N., Macura, T. Eckley, D.M., Johnston, J., Goldberg, I.G., 2008, {\it BMC Source Code for Biology and Medicine}, 3, 13

\bibitem[Shamir(2009)]{Sha09}
Shamir, L., 2009, \mnras, 399, 1367

\bibitem[Shimasaku et al.(1998)]{Shi01}
Shimasaku, K., et al. 2001, \aj, 122, 1238

\bibitem[Thorsten(2008)]{Tho08}
Thorsten, L., 2008, \apjs, 179, 319

\bibitem[Tyson(2002)]{Tys02}
Tyson, J.A., 2002, {\it Proceedings of the SPIE}, 4836, 10

\bibitem[York et al.(2000)]{Yor00}
York, D.G. et al., 2000, \aj, 120, 1579
%
\end{thebibliography}
\end{document}